\documentclass[conference]{IEEEtran}
\IEEEoverridecommandlockouts

\usepackage{multirow}
\usepackage[table,xcdraw]{xcolor}

\usepackage{etoolbox}
\makeatletter
\patchcmd{\@makecaption}
  {\scshape}
  {}  
  {}
  {}
\makeatother

\usepackage{amsmath} 

\usepackage{graphicx}

\usepackage{cite}
\usepackage{amsmath,amssymb,amsfonts}
\usepackage{algorithmic}
\usepackage{graphicx}
\usepackage{textcomp}
\usepackage{xcolor}
\def\BibTeX{{\rm B\kern-.05em{\sc i\kern-.025em b}\kern-.08em
    T\kern-.1667em\lower.7ex\hbox{E}\kern-.125emX}}

\begin{document}

\title{VM-DDPM: Vision Mamba Diffusion \\for Medical Image Synthesis }

\author{
\IEEEauthorblockN{Zhihan Ju}
\IEEEauthorblockA{Beijing University of \\Posts and Telecommunications\\juzhihan@bupt.edu.cn}
\and
\IEEEauthorblockN{Wanting Zhou}
\IEEEauthorblockA{Beijing University of\\Posts and Telecommunications\\wanting.zhou@bupt.edu.cn}
}

\maketitle

\begin{abstract}
In the realm of smart healthcare, researchers enhance the scale and diversity of medical datasets through medical image synthesis. However, existing methods are limited by CNN local perception and Transformer quadratic complexity, making it difficult to balance structural texture consistency. To this end, we propose the Vision Mamba DDPM (VM-DDPM) based on State Space Model (SSM), fully combining CNN local perception and SSM global modeling capabilities, while maintaining linear computational complexity. Specifically, we designed a multi-level feature extraction module called Multi-level State Space Block (MSSBlock), and a basic unit of encoder-decoder structure called State Space Layer (SSLayer) for medical pathological images. Besides, we designed a simple, Plug-and-Play, zero-parameter Sequence Regeneration strategy for the Cross-Scan Module (CSM), which enabled the S6 module to fully perceive the spatial features of the 2D image and stimulate the generalization potential of the model. To our best knowledge, this is the first medical image synthesis model based on the SSM-CNN hybrid architecture. Our experimental evaluation on three datasets of different scales, i.e.,  ACDC, BraTS2018, and ChestXRay, as well as qualitative evaluation by radiologists, demonstrate that VM-DDPM achieves state-of-the-art performance. 
\end{abstract}

\IEEEpeerreviewmaketitle

\section{Introduction}
With the development of smart healthcare, Data-driven Deep Learning technology\cite{1} has gained favor and attention from medical and computer experts due to its powerful nonlinear modeling ability and generalization advantages, and has emerged a large number of researches and applications in clinical diagnosis. This method can effectively assist professional doctors in clinical work such as pathological diagnosis\cite{2.1,2.2,2.3,2.4}, lesion segmentation\cite{3.1,3.2,3.3}, cross-modal synthesis\cite{4.1,4.2,4.3}, and diagnostic report writing\cite{5.1,5.2,5.3}, effectively improving consultation efficiency and diagnostic accuracy. On the contrary to its advantages, its training process relies heavily on large-scale and representative medical datasets. Limited by the rarity and scarcity of medical data, the high collection cost, and the privacy, security, and ethical issues contained therein, the practical application of such methods in real clinical environments still faces huge challenges. In order to solve this data bottleneck, more and more researchers have begun to focus on the research of unconditional medical image synthesis technology\cite{6.1,6.2,6.3}. This technology does not require any conditional information, and the generative model will learn the data distribution characteristics of real images and synthesize medical images from its potential variable space. This method can fundamentally improve the scale and diversity of rare medical image datasets, solve the privacy and security issues brought about by medical shared data, and serve physician education and training as well as downstream model training.
\par In recent years, research on unconditional medical image synthesis technology has developed rapidly. This technology mainly relies on two frameworks: Generative Adversarial Network (GAN)\cite{7} and Denoising Diffusion Probabilistic Model (DDPM)\cite{8}. The GAN consists of two parts: the Generator and the Discriminator. The Generator is responsible for modeling the potential distribution of the data and generating new samples, while the Discriminator is responsible for judging the authenticity of the samples. Through the game between the two, it is trained to ultimately achieve Nash Equilibrium. The method based on the GAN has achieved significant results in the diversity and quality of synthesized images\cite{9.1,9.2,9.3}, but it is prone to "mode collapse" or "layer disappearance" problems\cite{10} during the training process, causing training instability. In order to fundamentally solve the limitations of GANs, Ho et al.\cite{8} proposed DDPM based on the non-equilibrium thermodynamic diffusion model designed by Sohl-Dickstein et al.\cite{11}. The model gradually adds noise to the original image during the training process until it becomes a Gaussian distribution, and then reconstructs the original data source distribution during the reverse diffusion process to generate the image. This technique can effectively overcome the limitations of GANs, while possessing excellent image generation quality and diversity\cite{12}. Nevertheless, the diffusion model still has shortcomings. The diffusion model based on CNN is limited by its local receptive field, which greatly hinders its ability to capture long-distance information, making it difficult to model a complete pathological structure. The diffusion model based on Transformer performs well in global modeling, but the multi-head self-attention mechanism introduces quadratic computational complexity in image size, resulting in high computational costs, especially for tasks that require intensive prediction such as medical image synthesis, which greatly reduces the scalability of the model. The defects of existing models urgently require us to propose a medical image synthesis solution that maintains linear computational complexity while ensuring global modeling capabilities.
\par Recently, in order to overcome the challenges brought by long sequence modeling, State Space Model (SSM) has received widespread attention. Based on classical SSM research\cite{13}, researchers proposed a modern SSM (Mamba)\cite{14} architecture based on selection mechanisms and hardware-aware algorithms. It has been verified in multiple dense data domain tests that it not only has global modeling capabilities comparable to Transformer structures, but also has linear computational efficiency. This Method has achieved rapid development in the field of Natural Language Processing. Meanwhile, in the research of Zhu et al.\cite{15}, the Vision Mamba was first proposed for image classification, combined with bidirectional SSM for global visual context modeling and position embedding, and the Cross-Scan Module (CSM) was designed to bridge the gap between 1-D array scanning and 2-D plane traversal, introducing Mamba into the field of Computer Vision. Subsequently, Ruan et al.\cite{16} first proposed a medical image segmentation UNet model based on pure Vision Mamba (VM-UNet), demonstrating the huge potential of Mamba in modeling complex image distributions. Due to the excellent generalization performance and computational efficiency of this architecture, we believe it has the potential to solve the global modeling and computational bottlenecks currently encountered by medical image synthesis, while maintaining higher scalability.
\par In this paper, inspired by the application of VMamba\cite{15} in image classification, we propose the Vision Mamba Denoising Diffusion Probabilistic Model (VM-DDPM) with a hybrid architecture of Mamba and CNN, which fully utilizes the local perception of CNN and the long-distance modeling capabilities of SSM while maintaining linear computational complexity, aiming to explore the application potential of Mamba in medical image synthesis. Specifically, we designed a multi-level feature extraction module with time embedding, called Multi-level State Space Block (MSSBlock). It integrates the advantages of CNN local feature perception and SSM global modeling, and extracts features from multiple levels on the same medical image, ultimately achieving multi-level feature fusion of structure, texture, and lesion, thereby improving the structural integrity and texture consistency of synthetic medical images. Based on MSSBlock, we propose the State Space Layer and use it as the basic unit to conduct the complete encoder-decoder architecture, realizing the forward and backward propagation process of the diffusion model, and extending Mamba to medical image synthesis. In addition, we have improved the CSM mechanism and proposed the Sequence Regeneration strategy, which rearranges the images at the patch level before S6 operation, shuffles the order of the 2D space, quickly adds additional differential scanning sequences during the CSM process, enhances the Mamba module's ability to pay attention to the space continuity of medical images, and maximizes the feature extraction potential of the model.
\par Our contributions could be summarized as follows.
\begin{itemize}
\item We propose the Vision Mamba Denoising Diffusion Probabilistic Model (VM-DDPM), marking the first occasion of exploring the applications of SSM-CNN hybrid architecture in medical image synthesis.

\item We propose a multi-level feature extraction fusion module (MSSBlock) and a basic layer unit (SSLayer), constructing the encoder-decoder structure based on this.

\item We desige a simple, Plug and Play, zero-parameter Sequence Regeneration strategy to improve the original CSM mechanism and maximize the generalization potential for spatial continuity.

\item We conduct experiments on ACDC, BraTS2018, and ChestXRay datasets, as well as qualitative evaluation by radiologists, demonstrating that VM-DDPM has competitive performance and establishing a baseline for the application of SSM models in medical image synthesis. 
\end{itemize}

\section{Related work}
\textbf{Diffusion Models for Medical Image Synthesis.} In the past, medical image synthesis mainly relied on Generative Adversarial Network (GAN)\cite{7}, which consists of two parts: the Generator and the Discriminator, trained by the mutual game between the two. Common models include DCGAN\cite{17}, WGAN\cite{18}, SAGAN\cite{19}, PGGAN\cite{20}, UNetGAN\cite{21}, StytleGAN\cite{22}, etc. These networks have been proven to synthesize high-quality and diverse medical image samples. However, in a large number of studies\cite{10}, it has been found that due to its special training mechanism, the discriminator of GANs is prone to overfitting, pattern collapse or layer disappearance, causing serious training instability. This feature makes them unsuitable for synthesizing samples from rare conditions or imbalanced datasets, making it difficult to apply in the medical field. In recent years, the Denoising Diffusion Probabilistic Model (DDPM)\cite{8} has attracted a lot of attention from researchers, which consists of the forward Gaussian noise process and the reverse denoising process. Due to its high quality, wide pattern coverage, and training stability, researchers have begun to explore its application in the field of medical image synthesis. For example, in the field of histopathology, histological images of certain cancer subtypes are very rare. Moghadam et al.\cite{23} studied using DDPM to synthesize high fidelity histological images to achieve data amplification of rare images. In the field of thoracic and pulmonary diagnostics, Packhauser et al.\cite{24} synthesized high-quality quasi-conditional chest X-ray images through diffusion models to reduce the biometric information contained in the images and protect patient privacy. In addition, researchers have conducted extensive model structure exploration and improvement for specific tasks. For example, Aversa et al.\cite{25} proposed a parallel hierarchical diffusion model called DiffInfinite based on DDPM for large-scale medical image synthesis. Peng et al.\cite{26} proposed LW-DDPM, which reduces the sampling cost of diffusion models and improves the scalability of models by designing lightweight attention modules. Pan et al.\cite{27} improved the diffusion model structure based on the Swin-transformer network, enhancing its global modeling ability for structural textures in medical images. Compared with traditional models, it achieved better evaluation results and visualization effects. The DDPMs have been widely used in the field of medical image synthesis due to generalization and stability, but they still face the problem of global modeling and computational cost. This paper introduces the new architecture of Mamba\cite{14}, which is expected to solve this problem and make the methods have stronger application prospects.

\par \textbf{State Space Models for long sequence modeling.} Recently, researchers proposed the Selective State Space Model (Mamba)\cite{14} based on the Structured State Space Model (SSM)\cite{13}. Through the design of SSM variable parameters and efficient hardware-aware algorithms, it has achieved sequence long-distance modeling capabilities comparable to Transformer in Natural Language Processing, while also possessing linear computational complexity. Given the huge application potential of Mamba and the limitations of global perception ability and computational complexity under CNN and Transformer architectures, Zhu et al.\cite{15} proposed a new universal visual backbone (Vim) with bidirectional Mamba blocks, and achieved global contextual feature representation of images through position embedding and bidirectional scanning design, which proved its effectiveness in visual tasks. Subsequently, Ma et al.\cite{28} proposed the U-Mamba, which first applied the SSM-CNN hybrid model to medical image segmentation, combined the local feature extraction ability of the convolution layer with the ability of SSM to capture long-term dependencies, and achieved SOTA performance on multiple medical datasets. Meanwhile, Xing et al.\cite{29} also proposed an SSM-CNN hybrid model called SegMamba for 3D brain tumor segmentation, which attempted to combine SSM in the encoder and still use CNN in the decoder, achieving great performance. In addition, Ruan et al.\cite{16} proposed a medical image segmentation model VM-UNet constructed purely by SSM, which consists of Visual State Space (VSS) blocks to form a complete encoder-decoder structure, achieving competitive performance compared to traditional models. In this paper, inspired by these works, we construct a denoising diffusion probabilistic model under the SSM-CNN hybrid architecture for the first time and explore its potential applications in medical image synthesis.

\begin{figure*}[htp]
    \centering
    \includegraphics[width=18.2cm]{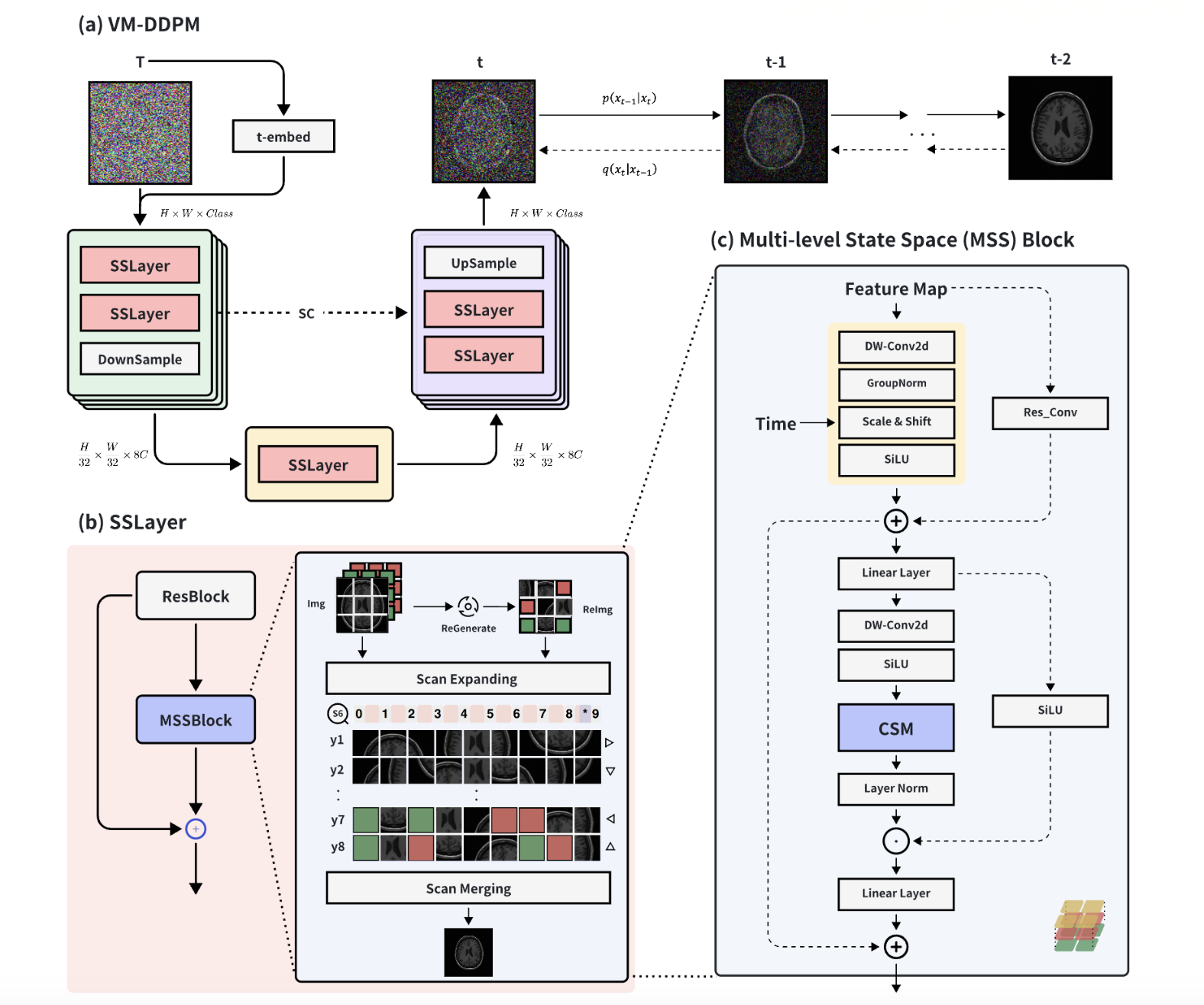}
    \caption{(a) The overall architecture of VM-DDPM. (b) SSLayer is the main construction layer of VM-DDPM. (c) MSSBlock is the core component of SSLayer, which includes time embedding and CSM operations.}
    \label{fig:1}
\end{figure*}

\section{Methods}
Firstly, in section 4.1, we explained the basic principles of the Denoising Diffusion Probabilistic Model (DDPM) and the State Space Model (SSM). Then, in section 4.2, we introduced the overall framework of the VM-DDPM, the feature extraction process, and the basic layer structure SSLayer that makes up its encoder-decoder architecture. In section 4.3, we focused on the core component unit of the model, MSSBlock, and introduced the principles and motivations of its multi-level design. Finally, in section 4.4, we detailed the core mechanism Cross-Scan Module (CSM) of Vision Mamba and the Sequence Regeneration strategy we designed for it.

\subsection{Preliminaries}
\textbf{Diffusion Models.} The Denoising Diffusion Probabilistic Model (DDPM)\cite{8} is an image generation model proposed based on the non-equilibrium thermodynamic diffusion model\cite{11}, which includes two steps: forward denoising process and reverse denoising process, as shown in Figure 1 (a). Among them, the forward process is defined as a Markov chain, which gradually adds noise to the original image in successive steps to obtain noise samples that conform to the Gaussian distribution. The forward transition probability can be expressed as follows:

\begin{equation}
q(x_{t}|x_{t-1}) = N(x_{t};\sqrt{1-\beta_{t}}\cdot x_{t-1}, \beta_{t}\cdot I), \forall{t} \in \lbrace 1,\ldots,T \rbrace
\end{equation}

Among them, $T$ and $\beta_{1}, \ldots, \beta_{T} \in [0,1)$ represent the number of diffusion steps and the variance scheduling during diffusion steps, respectively. $I$ is the identity matrix and $N(x;\mu,\sigma)$ represents the normal distribution of mean $\mu$ and covariance $\sigma$. The reverse process also follows the Markov chain from $x_{t}$ to $x_{0}$, 
gradually reducing the noise sample to the real image, which can be expressed as follows:

\begin{equation}
p(x_{t-1}|x_{t}) = N(x_{t-1};\mu(x_{t},t),\sum (x_{t},t))
\end{equation}

To train this model such that $p(x_{0})$ learns the true data distribution $q(x_{0})$, we can optimize the following variational bound on negative log-likelihood:

\begin{equation}
L_{vb}=E_{q(x_{0:T})}\lbrack \log \cfrac{p_{\theta}(x_{0:T})}{q(x_{1:T}|x_{0})} \rbrack \leq E_{q(x_{0})} \lbrack \log p_{\theta}(x_{0}) \rbrack
\end{equation}

Some work\cite{8} directly train the model $\epsilon_{\theta}(x_{t},t) $ to predict $\epsilon$, thereby re-parameterizing the equation and adopting a simplified goal.

\begin{equation}
L_{simple}=E_{t, x_{0}, \epsilon}\lbrack ||\epsilon - \epsilon_{\theta}(x_{t},t)||^2 \rbrack
\end{equation}

\par \textbf{SSM-Based Models.} Both structured state space models (S4)\cite{13} and Mamba\cite{14} rely on a classical continuous system that maps a one-dimensional input function or sequence (represented as $x(t) \in R$) to an output $y(t) \in R$ through intermediate implicit states ($h(t) \in R^N$). State space models have been shown to handle long-range dependencies theoretically and empirically by linearly scaling the $w.r.t$ sequence length. In their general form, they can be written in the form of linear ordinary differential equations (ODEs) as follows:

\begin{equation}
\begin{aligned}
h^{\prime} (t) = A(t)h(t) + B(t)x(t) \\
y(t) = C(t)h(t) + D(t)x(t)
\end{aligned}
\end{equation}

Among them, the parameters $A$, $B$, $C$, and $D$ of each layer can be learned through layer descent. S4 and Mamba introduce an additional time scale parameter $\Delta$, and use the zero-order preservation (ZOH) rule to discretize the parameters, making them more suitable for deep learning scenarios. The process is defined as follows:

\begin{equation}
\begin{aligned}
&\overline{A} = \exp({\Delta A}) \\
&\overline{B} = (\Delta A)^{-1}(\exp{\Delta A} - I) \cdot \Delta B
\end{aligned}
\end{equation}

After discretization, SSM-based models can be computed through linear recursion or global convolution.

\begin{equation}
\begin{aligned}
&\overline{K} = (C\overline{B},C\overline{AB},\ldots,C\overline{A}^{L-1}\overline{B}) \\
&y=x * \overline{K}
\end{aligned}
\end{equation}

Among them, $\overline{K}\in R^L$ represents the structured convolution kernel, and $L$ represents the length of the input sequence $x$. Mamba greatly improves the flexibility of SSM by relaxing the time-invariant constraints of SSM parameters while maintaining computational efficiency.

\subsection{Vision Mamba DDPM (VM-DDPM)}
Vision Mamba Denoising Diffusion Probabilistic Models (VM-DDPM) is built on a SSM-CNN hybrid architecture, which well integrates the local perception ability of the convolution layer and the global modeling ability of Mamba, while maintaining linear computational complexity, with higher scalability and application potential.
\par The model structure of VM-DDPM is shown in Figure 1 (a), which follows the encoder-bottleneck-decoder network structure, and each part supports vectorized embedding of time information. The encoder extracts and reduces the pathological features in the medical image layer by layer, and finally outputs the encoding vector in the high-dimensional feature space. The Bottleneck performs additional feature extraction and adjustment on the feature vector. The decoder reduces and upgrades the feature vector layer by layer until a denoising result of the same size as the original image is obtained. In addition, we refer to the classic medical image segmentation model UNet, add skip connections between the encoder and decoder, and perform feature fusion on the feature map extracted from the encoding path and the corresponding resolution on the decoding path to improve the feature positioning accuracy and training stability of the model.
\par Specifically, the encoder and decoder have a symmetrical structure, both consisting of four sampling stages. Each stage performs up or down sampling operations to increase or decrease the size of input features and adjust the number of channels. For each stage, we use the same hierarchical configuration, which is to connect two State Space Layers (SSLayers) in series and perform up or down sampling operations. Each SSLayer, as shown in Figure 1 (b), consists of a ResNet block that supports time embedding and a Multi-level State Space Block (MSSBlock). At the same time, residual connections and feature fusion within the layer are performed in the output connection to fully leverage the advantages of CNN local perception and SSM global modeling. Assuming the input image size is (B, C, H, W), after passing through the encoder, it will output feature vectors of (B, 8C, H/32, W/32). After passing through one layer of SSLayer, the feature vector is input to the decoder. Before each stage of upsampling, it is spliced and fused with the encoded feature map in the skip connection, and after the last stage, the feature dimension is restored through the final projection layer, and the denoising result of the original image size is finally output.

\subsection{Multi-Scale State Space Block (MSSBlock)}
Multi-Scale State Space Block is the core unit of SSLayer, as shown in Figure 1 (c). It fully combines the advantages of deep convolution operators and CSM operators, while learning local and global context features. Each MSSBlock is organized as a cascade of CSM modules that extract remote context features and CNN modules that extract local features of input feature maps, while achieving vectorized embedding of time information through the convolution layer.
\par Due to the fact that medical images often have dense feature distributions of structure, texture, lesions, and organs at multiple levels and scales, we inserted multiple external skip connections around the CSM module and CNN module to create multiple information transmission paths flowing through different modules. These paths transmit multiple sets of features at different levels to the output: (a) input features from the previous network; (b) local perceptual features calculated by the CNN module; (c) long-distance global modeling features calculated by the CSM module; (d) mixed local-contextual features processed by both the CNN and CSM modules, ultimately achieving multi-level feature fusion in the output.

\subsection{CSM with Sequence Regeneration}
Drawing on the work of Vision Mamba\cite{15}, the Cross-Scan Module (CSM) mainly consists of three parts: scan expansion operation, S6 block, and scan merge operation, as shown in Figure 1 (b). Among them, we have improved the scan expansion operation and proposed the Sequence Regeneration strategy based on the understanding of image continuity. The original scan expansion operation was designed based on the idea of bidirectional SSM, which expanded the two-dimensional input image at the patch level from the upper left corner or lower right corner in four directions: up, down, left, and right, thus forming four one-dimensional vector sequences. In this way, during the S6 operation, any area of the image will integrate information from other areas in different directions, and finally merge the information in different directions to form a complete new sequence and merge it into image form.
\par The design of the CSM mechanism can ensure that the S6 block scans the image from four different directions, thereby capturing the spatial features in the 2D image. However, existing scanning methods may not be sufficient for medical image synthesis, as they cannot guarantee that the S6 module evenly pays attention to the contextual relationship between each patch and the rest of the patches. Any attention bias generated will seriously affect the spatial continuity and texture consistency of the synthesized medical image.
\par Based on this, we propose a simple, Plug and Play, zero-parameter CSM module improvement method, the Sequence Regeneration strategy. Before the scan expansion operation, the input image is randomly shuffled in the patch dimension to regenerate a brand new patch combination result, thereby obtaining a brand new position vector sequence during the scan expansion process. It should be noted that this method randomly generates a sequence order before each S6 operation. When the model is fully trained, the S6 module will receive a diverse and evenly distributed set of extended sequences, thus maintaining sufficient attention to each patch context feature without adding any additional parameters and maintaining linear computational efficiency. This method will effectively improve the generalization performance of the SSM model in medical image synthesis.

\begin{table*}[]
\renewcommand\arraystretch{1.6}
\caption{Comparative experimental results on ChestXRay, BraTS2018 and ACDC datasets.}
\setlength{\tabcolsep}{6.9mm}{
\begin{tabular}{ccccccc}
\hline
\textbf{Models} & \textbf{Dataset A}                                                                & \multicolumn{1}{c}{\textbf{FID$\downarrow$}} & \textbf{Dataset B}                                                               & \multicolumn{1}{c}{\textbf{FID$\downarrow$}} & \multicolumn{1}{l}{\textbf{Dataset C}}                                      & \multicolumn{1}{c}{\textbf{FID$\downarrow$}} \\ \hline
DCGAN\cite{33}           & \multirow{6}{*}{\begin{tabular}[c]{@{}c@{}}ChestXRay\\ (34999 imgs)\end{tabular}} & 98.216                           & \multirow{6}{*}{\begin{tabular}[c]{@{}c@{}}BraTS2018\\ (6528 imgs)\end{tabular}} & 83.768                           & \multirow{6}{*}{\begin{tabular}[c]{@{}c@{}}ACDC\\ (1798 imgs)\end{tabular}} & 89.473                           \\
SAGAN\cite{19}           &                                                                                   & 42.753                           &                                                                                  & 90.582                           &                                                                             & 79.037                           \\
UNetGAN\cite{21}         &                                                                                   & 94.668                           &                                                                                  & 60.097                           &                                                                             & 82.113                           \\
DDPM\cite{23}            &                                                                                   & 23.679                           &                                                                                  & 28.335                           &                                                                             & 66.630                           \\
DDIM\cite{34}            &                                                                                   & 21.695                           &                                                                                  & 22.391                           &                                                                             & 47.434                           \\
\textbf{VM-DDPM}         &                                                                                   & \textbf{11.783}                           &                                                                                  & \textbf{12.513}                           &                                                                             & \textbf{34.525}                           \\ \hline
\end{tabular}}
\end{table*}

\begin{table*}[]
\renewcommand\arraystretch{1.6}
\caption{Ablation studies on Sequence Regenerate strategy of VM-DDPM.}
\setlength{\tabcolsep}{5.4mm}{
\begin{tabular}{ccccccc}
\hline
\multicolumn{1}{c}{\textbf{Modules}} & \textbf{Dataset A}                                                                & \multicolumn{1}{c}{\textbf{FID$\downarrow$}} & \textbf{Dataset B}                                                               & \multicolumn{1}{c}{\textbf{FID$\downarrow$}} & \multicolumn{1}{l}{\textbf{Dataset C}}                                      & \multicolumn{1}{c}{\textbf{FID$\downarrow$}} \\ \hline
VM-DDPM + CSM                        & \multirow{2}{*}{\begin{tabular}[c]{@{}c@{}}ChestXRay\\ (34999 imgs)\end{tabular}} & 16.783                           & \multirow{2}{*}{\begin{tabular}[c]{@{}c@{}}BraTS2018\\ (6528 imgs)\end{tabular}} & 13.578                           & \multirow{2}{*}{\begin{tabular}[c]{@{}c@{}}ACDC\\ (1798 imgs)\end{tabular}} & 40.691                           \\
\textbf{VM-DDPM + Regenerate CSM}             &                                                                                   & \textbf{11.783}                           &                                                                                  & \textbf{12.513}                           &                                                                             & \textbf{34.525}                           \\ \hline
\end{tabular}}
\end{table*}

\section{Experiment}

\subsection{Experimental Setup}
\textbf{Datasets.} In this paper, we conducted experiments for VM-DDPM and its baselines on three medical public datasets of different scales and locations: ACDC\cite{30}, BraTS2018\cite{31}, and ChestXRay\cite{32}. ACDC is a public dataset containing 1798 2D cardiac MRIs images from 3DMR scans in the Automated Cardiac Diagnosis Challenge (ACDC). The MR scans were collected as a series of axial slices covering the heart base to the left ventricle, with an axial line spacing of 5 to 8 millimeters (mm). The 2D axial slices were then separated to form a training dataset. BraTS2018 is a 3D brain MRI dataset from 285 patients, with four modalities for each case. In this paper, we use 6528 2D brain images under a single modality for model training and testing. ChestXRay is a public dataset containing 112,120 2D images of chest X-rays. We use the first 34,999 images for experiments, maintaining a three-fold scale increase compared to the ACDC and BraTS2018 datasets.

\par \textbf{Evaluation Metrics.} In this paper, we use Fred'chet Inception Distance (FID) as the main metric. FID is a comprehensive metric that has been proven to be more consistent with human evaluation in evaluating the realism and variation of synthetic images. In all of our experiments, FID was calculated using 50,000 synthetic images to obtain objective and fair evaluation results.

\par \textbf{Baselines.} In this paper, in order to demonstrate the advantages of VM-DDPM as a denoising diffusion model compared to traditional generative adversarial networks and the advantages of Vision Mamba architecture compared to traditional CNNs, we selected three classic GANs and two classic DDPMs as baselines for horizontal comparison, namely DCGAN\cite{33}, SAGAN\cite{19}, UNetGAN\cite{21}, DDPM\cite{23}, and DDIM\cite{34}. Most of these models have been applied in the field of medical image synthesis, and we chose their expansion methods in medical tasks as the baseline.

\par \textbf{Implementation.} All methods involved in this experiment are implemented with reference to the original paper or open source code on the internet. Due to limited computing resources and dataset size, during the experiment, we uniformly scaled the image center to 128 × 128 resolution and performed regular normalization, image flipping and other preprocessing operations to avoid model overfitting. In addition, the DDPM-related model loss function uses Mean Square Error Loss, and the GAN-related model loss function uses Binary Cross Entropy Loss. All models use Adam optimizer, and uniformly distributed noise vectors $z \in [-1, 1]^{1000}$ are used as input in model training and testing, with a minimum Learning Rate of $1e-4$ and a batch size of 16. All experiments were conducted on a single NVIDIA GeForce RTX 4090 GPU with a memory size of 24GB.

\begin{figure*}[htp]
    \centering
    \includegraphics[width=18cm]{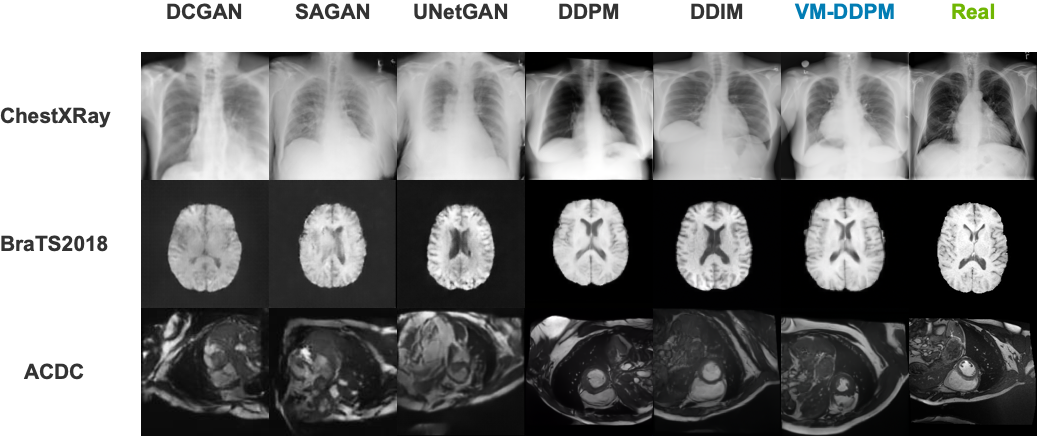}
    \caption{Medical image synthesis samples and the real images.}
    \label{fig:2}
\end{figure*}

\subsection{Main Results}
We selected three publicly available medical image datasets from different locations, and the dataset size tripled. Based on this, we compared VM-DDPM with the most advanced models for medical image synthesis. Among them, we demonstrated the performance advantages of the DDPM framework by comparing VM-DDPM with GAN-based models, and demonstrated the improvement in generalization ability of the SSM-CNN hybrid architecture compared to the pure CNN architecture by comparing VM-DDPM with DDPM-based methods. The comparison results of FID refer to Table 1, and the visualization results are shown in Figure 2. In the testing of the three datasets, the FID of VM-DDPM was significantly lower than that of the GAN-related models, while outperforming the existing DDPM models. Although this performance advantage is gradually narrowing on the small-scale dataset ACDC, it still achieves the best performance. The results on medical datasets of different sizes in different parts prove that VM-DDPM has great generalization ability and wide pattern coverage. At the same time, based on the visualization results, the significant improvement in global modeling ability of VM-DDPM can be intuitively felt, and its synthesized images have more realistic pathological structures and detailed textures.

\subsection{Ablation Studies}
In this section, we use three datasets to perform ablation studies on the core component structure Cross-Scan Module (CSM) of VM-DDPM. We compare the impact of the original CSM and the CSM combined with the Sequence Regeneration strategy on the model performance. It can be found that the model using the Sequence Regeneration strategy has a certain degree of performance improvement on the three datasets, proving the effectiveness of the strategy. In addition, we find that the improvement on the BraTS2018 dataset is slightly smaller than the other two datasets. We speculate that the improvement in the model's spatial feature extraction ability has a small impact on the overall synthesis due to the large irrelevant parts of the images on the dataset, which indirectly confirms the theoretical effectiveness of the strategy in spatial continuity.

\begin{table}[]
\renewcommand\arraystretch{1.6}
\caption{The Turing test results of Doctor 1.}
\setlength{\tabcolsep}{3.0mm}{
\begin{tabular}{ccccccc}
\hline
\textbf{Doctor 1} & \textbf{Judge Real} & \textbf{Judge Syn.} & \textbf{Precision} & \textbf{Recall} \\ \hline
Data Real         & 86                  & 14                  & 0.860              & 0.515           \\
Data Syn.         & 81                  & 17                  & 0.173              & 0.548           \\ \hline
\end{tabular}}
\end{table}

\begin{table}[]
\renewcommand\arraystretch{1.6}
\caption{The Turing test results of Doctor 2.}
\setlength{\tabcolsep}{3.0mm}{
\begin{tabular}{ccccccc}
\hline
\textbf{Doctor 2} & \textbf{Judge Real} & \textbf{Judge Syn.} & \textbf{Precision} & \textbf{Recall} \\ \hline
Data Real         & 44                  & 28                  & 0.611              & 0.431           \\
Data Syn.         & 58                  & 21                  & 0.266              & 0.429           \\ \hline
\end{tabular}}
\end{table}

\begin{table}[]
\renewcommand\arraystretch{1.6}
\caption{The bilateral Fisher's precise test results.}
\setlength{\tabcolsep}{6.5mm}{
\begin{tabular}{ccccccc}
\hline
\textbf{}              & \textbf{Doctor 1} & \textbf{Doctor 2} \\ \hline
\textbf{Fisher-exact’s p-value} & 0.56136           & 0.11982           \\ \hline
\end{tabular}}
\end{table}

\subsection{Qualitative Evaluation by Radiologist}
In this section, we invited two radiologists with different levels of expertise to assist us in conducting a Turing test\cite{35} to fully compare the morphological characteristics of synthetic images and real images. Specifically, we randomly selected one hundred synthetic images generated by VM-DDPM trained on the BraTS2018 dataset, as well as one hundred real images in the dataset. At the same time, we designed a pathological survey consisting of the following two questions for the two experts. The first question asked participants whether they believed the images were real or synthetic, and the second question asked them about their Confidence Level (divided into low, medium, and high levels) for their own judgment. The first radiology expert (Doctor 1) is an attending physician at Beijing Tiantan Hospital, and the second (Doctor 2) is a doctoral student at the School of Medicine. The test results are shown in Tables 3 and 4. For the judgment results of medium to high Confidence Level, we calculated the Precision and Recall metrics under real and synthetic images, and found that the doctors' judgments on the two were similar. This result shows that all participating experts were unable to distinguish the morphological differences between the synthetic images generated by VM-DDPM and the real images. For most of synthetic images correctly identified by experts, their confidence level was low. Our investigation results show that the medical synthetic images generated by VM-DDPM are very similar to real images, with almost identical pathological features and detailed textures, which can fully serve the purpose of rare data augmentation and physician training. In addition, we used bilateral Fisher's precise test to verify whether there were differences between the observations of the two experts on real images and synthetic images, as shown in Table 5. The results showed that there was no statistically significant difference in their performance in identifying the authenticity of images, further verifying the above investigation conclusions.

\section{Conclusion}
In this paper, we introduce the Vision Mamba Denoising Diffusion Probabilistic Model (VM-DDPM) based on the SSM-CNN hybrid architecture for the first time, establishing a baseline for the application of SSM models in medical image synthesis. In order to combine the advantages of CNN local perception and SSM global modeling, we design a multi-scale feature fusion module (MSSBlock) and its corresponding layer structure (SSLayer), and construct a complete encoder-decoder structure based on this. In addition, we propose a simple, Plug-and-Play, zero-parameter Sequence Regeneration strategy for the CSM mechanism, aiming to fully tap the generalization potential of the S6 module for 2D space features. VM-DDPM was experimented on three public medical datasets and tested by professional radiologists. The results show that the model has strong competitiveness in medical image synthesis and has huge application space.

\bibliographystyle{IEEEtran}

\bibliography{related_work}

\begin{thebibliography}{10}
\providecommand{\url}[1]{#1}
\csname url@samestyle\endcsname
\providecommand{\newblock}{\relax}
\providecommand{\bibinfo}[2]{#2}
\providecommand{\BIBentrySTDinterwordspacing}{\spaceskip=0pt\relax}
\providecommand{\BIBentryALTinterwordstretchfactor}{4}
\providecommand{\BIBentryALTinterwordspacing}{\spaceskip=\fontdimen2\font plus
\BIBentryALTinterwordstretchfactor\fontdimen3\font minus \fontdimen4\font\relax}
\providecommand{\BIBforeignlanguage}[2]{{%
\expandafter\ifx\csname l@#1\endcsname\relax
\typeout{** WARNING: IEEEtran.bst: No hyphenation pattern has been}%
\typeout{** loaded for the language `#1'. Using the pattern for}%
\typeout{** the default language instead.}%
\else
\language=\csname l@#1\endcsname
\fi
#2}}
\providecommand{\BIBdecl}{\relax}
\BIBdecl

\bibitem{1}
A.~Maier, C.~Syben, T.~Lasser, and C.~Riess, ``A gentle introduction to deep learning in medical image processing,'' \emph{Zeitschrift f{\"u}r Medizinische Physik}, vol.~29, no.~2, pp. 86--101, 2019.

\bibitem{2.1}
J.~Ker, L.~Wang, J.~Rao, and T.~Lim, ``Deep learning applications in medical image analysis,'' \emph{Ieee Access}, vol.~6, pp. 9375--9389, 2017.

\bibitem{2.2}
S.~S. Yadav and S.~M. Jadhav, ``Deep convolutional neural network based medical image classification for disease diagnosis,'' \emph{Journal of Big data}, vol.~6, no.~1, pp. 1--18, 2019.

\bibitem{2.3}
S.~Azizi, B.~Mustafa, F.~Ryan, Z.~Beaver, J.~Freyberg, J.~Deaton, A.~Loh, A.~Karthikesalingam, S.~Kornblith, T.~Chen \emph{et~al.}, ``Big self-supervised models advance medical image classification,'' in \emph{Proceedings of the IEEE/CVF international conference on computer vision}, 2021, pp. 3478--3488.

\bibitem{2.4}
Y.~Dai, Y.~Gao, and F.~Liu, ``Transmed: Transformers advance multi-modal medical image classification,'' \emph{Diagnostics}, vol.~11, no.~8, p. 1384, 2021.

\bibitem{3.1}
P.~Malhotra, S.~Gupta, D.~Koundal, A.~Zaguia, W.~Enbeyle \emph{et~al.}, ``Deep neural networks for medical image segmentation,'' \emph{Journal of Healthcare Engineering}, vol. 2022, 2022.

\bibitem{3.2}
X.-X. Yin, L.~Sun, Y.~Fu, R.~Lu, Y.~Zhang \emph{et~al.}, ``U-net-based medical image segmentation,'' \emph{Journal of healthcare engineering}, vol. 2022, 2022.

\bibitem{3.3}
H.~Cao, Y.~Wang, J.~Chen, D.~Jiang, X.~Zhang, Q.~Tian, and M.~Wang, ``Swin-unet: Unet-like pure transformer for medical image segmentation,'' in \emph{European conference on computer vision}.\hskip 1em plus 0.5em minus 0.4em\relax Springer, 2022, pp. 205--218.

\bibitem{4.1}
O.~Dalmaz, M.~Yurt, and T.~{\c{C}}ukur, ``Resvit: residual vision transformers for multimodal medical image synthesis,'' \emph{IEEE Transactions on Medical Imaging}, vol.~41, no.~10, pp. 2598--2614, 2022.

\bibitem{4.2}
T.~Zhou, H.~Fu, G.~Chen, J.~Shen, and L.~Shao, ``Hi-net: hybrid-fusion network for multi-modal mr image synthesis,'' \emph{IEEE transactions on medical imaging}, vol.~39, no.~9, pp. 2772--2781, 2020.

\bibitem{4.3}
M.~{\"O}zbey, O.~Dalmaz, S.~U. Dar, H.~A. Bedel, {\c{S}}.~{\"O}zturk, A.~G{\"u}ng{\"o}r, and T.~{\c{C}}ukur, ``Unsupervised medical image translation with adversarial diffusion models,'' \emph{IEEE Transactions on Medical Imaging}, 2023.

\bibitem{5.1}
Z.~Chen, Y.~Shen, Y.~Song, and X.~Wan, ``Cross-modal memory networks for radiology report generation,'' \emph{arXiv preprint arXiv:2204.13258}, 2022.

\bibitem{5.2}
S.~Yang, X.~Wu, S.~Ge, Z.~Zheng, S.~K. Zhou, and L.~Xiao, ``Radiology report generation with a learned knowledge base and multi-modal alignment,'' \emph{Medical Image Analysis}, vol.~86, p. 102798, 2023.

\bibitem{5.3}
F.~Nooralahzadeh, N.~P. Gonzalez, T.~Frauenfelder, K.~Fujimoto, and M.~Krauthammer, ``Progressive transformer-based generation of radiology reports,'' \emph{arXiv preprint arXiv:2102.09777}, 2021.

\bibitem{6.1}
C.~Bermudez, A.~J. Plassard, L.~T. Davis, A.~T. Newton, S.~M. Resnick, and B.~A. Landman, ``Learning implicit brain mri manifolds with deep learning,'' in \emph{Medical imaging 2018: Image processing}, vol. 10574, 2018, pp. 408--414.

\bibitem{6.2}
C.~Baur, S.~Albarqouni, and N.~Navab, ``\BIBforeignlanguage{en-US}{Generating highly realistic images of skin lesions with gans.}'' \emph{\BIBforeignlanguage{en-US}{arXiv: Computer Vision and Pattern Recognition}}, Sep 2018.

\bibitem{6.3}
F.~Calimeri, A.~Marzullo, C.~Stamile, and G.~Terracina, ``Biomedical data augmentation using generative adversarial neural networks,'' in \emph{International conference on artificial neural networks}, 2017, pp. 626--634.

\bibitem{7}
I.~Goodfellow, J.~Pouget-Abadie, M.~Mirza, B.~Xu, D.~Warde-Farley, S.~Ozair, A.~Courville, and Y.~Bengio, ``Generative adversarial nets,'' \emph{Advances in neural information processing systems}, vol.~27, 2014.

\bibitem{8}
J.~Ho, A.~Jain, and P.~Abbeel, ``Denoising diffusion probabilistic models,'' \emph{Advances in neural information processing systems}, vol.~33, pp. 6840--6851, 2020.

\bibitem{9.1}
X.~Yi, E.~Walia, and P.~Babyn, ``Generative adversarial network in medical imaging: A review,'' \emph{Medical image analysis}, vol.~58, p. 101552, 2019.

\bibitem{9.2}
S.~Kazeminia, C.~Baur, A.~Kuijper, B.~van Ginneken, N.~Navab, S.~Albarqouni, and A.~Mukhopadhyay, ``Gans for medical image analysis,'' \emph{Artificial Intelligence in Medicine}, vol. 109, p. 101938, 2020.

\bibitem{9.3}
T.~Wang, Y.~Lei, Y.~Fu, W.~J. Curran, T.~Liu, and X.~Yang, ``Medical imaging synthesis using deep learning and its clinical applications: A review,'' \emph{arXiv preprint arXiv:2004.10322}, 2020.

\bibitem{10}
M.~Arjovsky, S.~Chintala, and L.~Bottou, ``Wasserstein generative adversarial networks,'' in \emph{International conference on machine learning}.\hskip 1em plus 0.5em minus 0.4em\relax PMLR, 2017, pp. 214--223.

\bibitem{11}
J.~Sohl-Dickstein, E.~Weiss, N.~Maheswaranathan, and S.~Ganguli, ``Deep unsupervised learning using nonequilibrium thermodynamics,'' in \emph{International conference on machine learning}.\hskip 1em plus 0.5em minus 0.4em\relax PMLR, 2015, pp. 2256--2265.

\bibitem{12}
P.~Dhariwal and A.~Nichol, ``Diffusion models beat gans on image synthesis,'' \emph{Advances in neural information processing systems}, vol.~34, pp. 8780--8794, 2021.

\bibitem{13}
R.~E. Kalman, ``A new approach to linear filtering and prediction problems,'' 1960.

\bibitem{14}
A.~Gu and T.~Dao, ``Mamba: Linear-time sequence modeling with selective state spaces,'' \emph{arXiv preprint arXiv:2312.00752}, 2023.

\bibitem{15}
Y.~Liu, Y.~Tian, Y.~Zhao, H.~Yu, L.~Xie, Y.~Wang, Q.~Ye, and Y.~Liu, ``Vmamba: Visual state space model,'' \emph{arXiv preprint arXiv:2401.10166}, 2024.

\bibitem{16}
J.~Ruan and S.~Xiang, ``Vm-unet: Vision mamba unet for medical image segmentation,'' \emph{arXiv preprint arXiv:2402.02491}, 2024.

\bibitem{17}
H.~Salehinejad, S.~Valaee, T.~Dowdell, E.~Colak, and J.~Barfett, ``Generalization of deep neural networks for chest pathology classification in x-rays using generative adversarial networks,'' in \emph{2018 IEEE international conference on acoustics, speech and signal processing (ICASSP)}.\hskip 1em plus 0.5em minus 0.4em\relax IEEE, 2018, pp. 990--994.

\bibitem{18}
C.~Han, H.~Hayashi, L.~Rundo, R.~Araki, W.~Shimoda, S.~Muramatsu, Y.~Furukawa, G.~Mauri, and H.~Nakayama, ``\BIBforeignlanguage{en-US}{Gan-based synthetic brain mr image generation},'' in \emph{\BIBforeignlanguage{en-US}{2018 IEEE 15th International Symposium on Biomedical Imaging (ISBI 2018)}}, 2018.

\bibitem{19}
H.~Zhang, I.~Goodfellow, D.~Metaxas, and A.~Odena, ``Self-attention generative adversarial networks,'' in \emph{International conference on machine learning}, 2019.

\bibitem{20}
A.~Beers, J.~Brown, K.~Chang, J.~P. Campbell, S.~Ostmo, M.~F. Chiang, and J.~Kalpathy-Cramer, ``High-resolution medical image synthesis using progressively grown generative adversarial networks,'' \emph{arXiv preprint arXiv:1805.03144}, 2018.

\bibitem{21}
E.~Schonfeld, B.~Schiele, and A.~Khoreva, ``A u-net based discriminator for generative adversarial networks,'' in \emph{Computer Vision and Pattern Recognition}, 2020, pp. 8207--8216.

\bibitem{22}
T.~Karras, S.~Laine, and T.~Aila, ``A style-based generator architecture for generative adversarial networks,'' in \emph{Proceedings of the IEEE/CVF conference on computer vision and pattern recognition}, 2019, pp. 4401--4410.

\bibitem{23}
P.~A. Moghadam, S.~Van~Dalen, K.~C. Martin, J.~Lennerz, S.~Yip, H.~Farahani, and A.~Bashashati, ``A morphology focused diffusion probabilistic model for synthesis of histopathology images,'' in \emph{Proceedings of the IEEE/CVF winter conference on applications of computer vision}, 2023, pp. 2000--2009.

\bibitem{24}
K.~Packh{\"a}user, L.~Folle, F.~Thamm, and A.~Maier, ``Generation of anonymous chest radiographs using latent diffusion models for training thoracic abnormality classification systems,'' in \emph{2023 IEEE 20th International Symposium on Biomedical Imaging (ISBI)}, pp. 1--5.

\bibitem{25}
M.~Aversa, G.~Nobis, M.~H{\"a}gele, K.~Standvoss, M.~Chirica, R.~Murray-Smith, A.~M. Alaa, L.~Ruff, D.~Ivanova, W.~Samek \emph{et~al.}, ``Diffinfinite: Large mask-image synthesis via parallel random patch diffusion in histopathology,'' \emph{Advances in Neural Information Processing Systems}, vol.~36, 2024.

\bibitem{26}
J.~Peng, G.~Chen, K.~Saruta, and Y.~Terata, ``2d brain mri image synthesis based on lightweight denoising diffusion probabilistic model,'' \emph{Medical Imaging Process \& Technology}, vol.~6, no.~1, 2023.

\bibitem{27}
S.~Pan, T.~Wang, R.~L. Qiu, M.~Axente, C.-W. Chang, J.~Peng, A.~B. Patel, J.~Shelton, S.~A. Patel, J.~Roper \emph{et~al.}, ``2d medical image synthesis using transformer-based denoising diffusion probabilistic model,'' \emph{Physics in Medicine \& Biology}, vol.~68, no.~10, p. 105004, 2023.

\bibitem{28}
J.~Ma, F.~Li, and B.~Wang, ``U-mamba: Enhancing long-range dependency for biomedical image segmentation,'' \emph{arXiv preprint arXiv:2401.04722}, 2024.

\bibitem{29}
Z.~Xing, T.~Ye, Y.~Yang, G.~Liu, and L.~Zhu, ``Segmamba: Long-range sequential modeling mamba for 3d medical image segmentation,'' \emph{arXiv preprint arXiv:2401.13560}, 2024.

\bibitem{33}
Y.~S. Devi and S.~P. Kumar, ``Dr-dcgan: A deep convolutional generative adversarial network (dc-gan) for diabetic retinopathy image synthesis,'' \emph{Webology (ISSN: 1735-188X)}, vol.~19, no.~2, 2022.

\bibitem{34}
J.~Song, C.~Meng, and S.~Ermon, ``Denoising diffusion implicit models,'' \emph{arXiv preprint arXiv:2010.02502}, 2020.

\bibitem{30}
C.~Sakaridis, D.~Dai, and L.~Van~Gool, ``Acdc: The adverse conditions dataset with correspondences for semantic driving scene understanding,'' in \emph{International Conference on Computer Vision}, 2021, pp. 10\,765--10\,775.

\bibitem{31}
B.~H. Menze, A.~Jakab, S.~Bauer, J.~Kalpathy-Cramer, K.~Farahani, J.~Kirby, Y.~Burren, N.~Porz, J.~Slotboom, R.~Wiest \emph{et~al.}, ``The multimodal brain tumor image segmentation benchmark (brats),'' \emph{IEEE transactions on medical imaging}, vol.~34, no.~10, 2014.

\bibitem{32}
X.~Wang, Y.~Peng, L.~Lu, Z.~Lu, M.~Bagheri, and R.~M. Summers, ``Chestx-ray8: Hospital-scale chest x-ray database and benchmarks on weakly-supervised classification and localization of common thorax diseases,'' in \emph{Proceedings of the IEEE conference on computer vision and pattern recognition}, 2017, pp. 2097--2106.

\bibitem{35}
A.~M. Turing, \emph{Computing machinery and intelligence}.\hskip 1em plus 0.5em minus 0.4em\relax Springer, 2009.

\end{thebibliography}

\end{document}